\documentclass[a4paper,11pt]{article}
\usepackage{aaskaiid}
\usepackage{orcidlink}
\usepackage{xeCJK}
\usepackage{zxjatype}
\usepackage[haranoaji]{zxjafont}

\title{VLBI Astrometry of Magnetars}
\ShortTitle{VLBI Astrometry of Magnetars}
\ShortName{Akahori et al.} 
\author[1,2]{Takuya Akahori (赤堀~卓也) \orcidlink{0000-0001-9399-5331}}
\emailAdd{takuya.akahori@nao.ac.jp}
\author[1,3]{Sujin Eie}
\emailAdd{astrosujin@gmail.com}
\author[1]{Kei Amada (甘田~渓) \orcidlink{0000-0003-4419-6132}}
\emailAdd{kei.amada@nao.ac.jp}
\author[1]{Sangita Kumari}
\emailAdd{sangita.kumari@nao.ac.jp}
\author[1,4]{Kohei Kurahara(藏原~昂平) \orcidlink{0000-0003-2955-1239}}
\emailAdd{kurahara.kohei.i7@f.mail.nagoya-u.ac.jp}
\author[1,5]{Hiroto Masaoka (正岡~滉翔) \orcidlink{0009-0003-5746-1510}}
\emailAdd{hmasaoka@g.ecc.u-tokyo.ac.jp}
\author[1]{Hao Ding}
\emailAdd{hidingastro@hotmail.com}
\author[6]{Teruaki Enoto}
\emailAdd{enoto.teruaki.2w@kyoto-u.ac.jp}
\author[7]{Shota Kisaka}
\emailAdd{kisaka@hiroshima-u.ac.jp}
\author[8]{Keiichi Maeda (前田~啓一) \orcidlink{0000-0003-2611-7269}}
\emailAdd{keiichi.maeda@kusastro.kyoto-u.ac.jp}
\author[9]{Kotaro Niinuma (新沼~浩太郎) \orcidlink{0000-0002-8169-3579}}
\emailAdd{niinuma@yamaguchi-u.ac.jp}
\affiliation[1]{Mizusawa VLBI Observatory, National Astronomical Observatory of Japan, 2-21-1 Mitaka, Tokyo 181-8588, Japan}
\affiliation[2]{SOKENDAI, Shonan Village, Hayama-machi, Miura-gun, Kanagawa 240-0193 Japan}
\affiliation[3]{Institute of Astronomy and Astrophysics, Academia Sinica, 11F of AS/NTU Astronomy-Mathematics Building, No.1, Section 4, Roosevelt Road, Taipei 10617, Taiwan}
\affiliation[4]{Kobayashi-Maskawa Institute for the Origin of Particles and the Universe (KMI), Nagoya University, Furo-cho, Chikusa-ku, Nagoya, Aichi 464-8601, Japan}
\affiliation[6]{Department of Astronomy, School of Science, The University of Tokyo, 7-3-1 Hongo, Bunkyo-ku, Tokyo, 113-0033, Japan}
\affiliation[7]{Department of Physics, Kyoto University, Kitashirakawa Oiwake, Sakyo, Kyoto 606-8502, Japan}
\affiliation[8]{Physics Program, Graduate School of Advanced Science and Engineering, Hiroshima University, Higashi-Hiroshima, Hiroshima 739-8526, Japan}
\affiliation[9]{Graduate School of Sciences and Technology for Innovation, Yamaguchi University, Yoshida 1677-1, Yamaguchi 753-8512, Japan}

\abstract{
The origin of the strongest magnetic fields in the Universe, i.e., the origin of magnetars, is a longstanding question. An enhanced dynamo effect in an irregular supernova explosion is a possible origin, which implies a stronger kick velocity and a higher proper motion of the magnetar compared to those of ordinary pulsars as well as an irregular morphology of the host supernova remnant (SNR). However, this hypothesis is not well studied yet, because there is a lack of precise measurement of the proper motion of magnetar and of identification of the host SNR. VLBI astrometry of magnetars is a unique tool to examine the hypothesis. In this chapter, we introduce the MONSTER (Monitoring Observations of the Neutron Stars That Evolve Rapidly) Project. SKA-VLBI's unprecedented sensitivity and the highest angular resolution will allow us to dramatically expand the survey volume in which we can measure the proper motion of magnetars within a radio outburst period of a few months.
}

\begin{document}
\maketitle

\section{Introduction}

\subsection{Magnetar Hypothesis and its Origin}

\begin{figure}[btp]
    \centering
	\includegraphics[width=\linewidth]{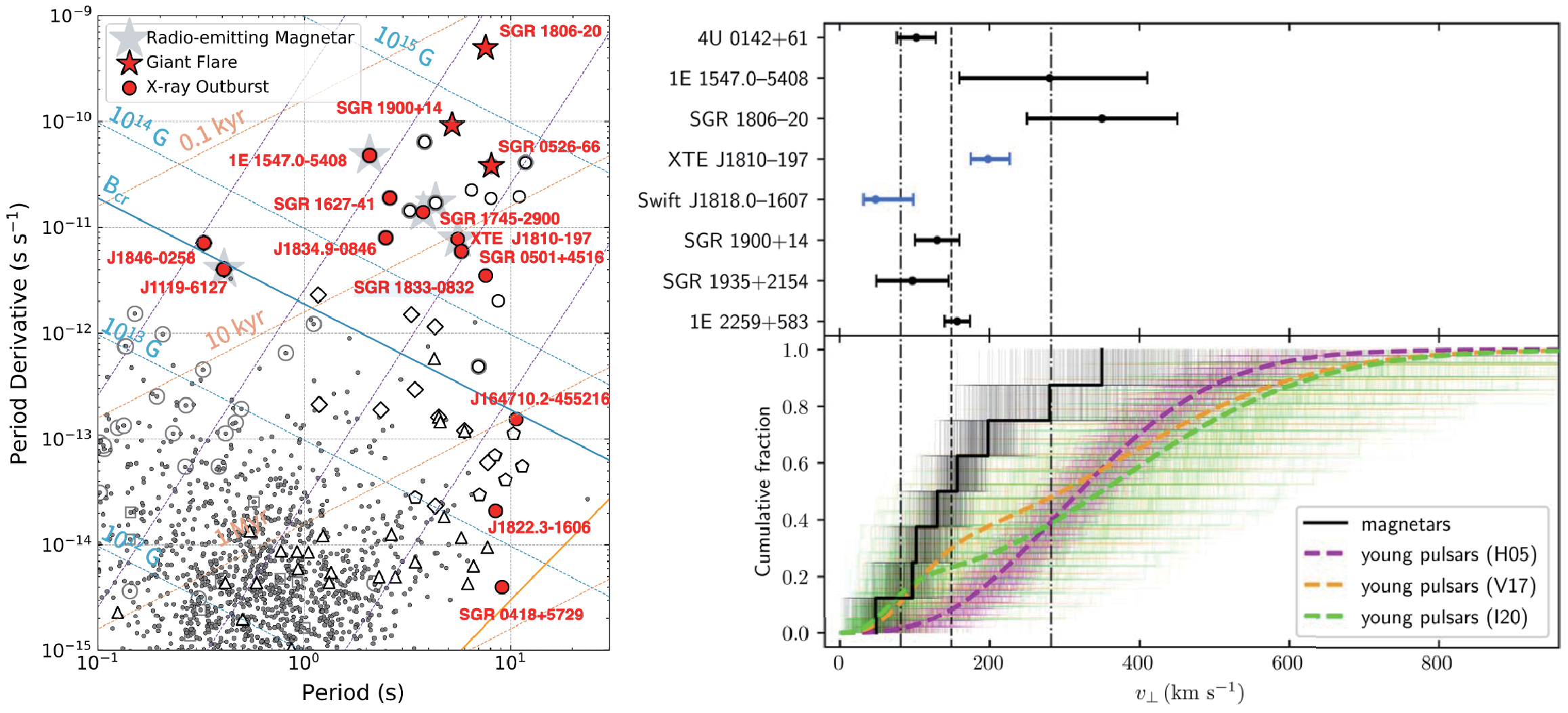}
    \caption{
    The left panel shows the $P-\dot{P}$ diagram \citep{2019RPPh...82j6901E}. The right panel shows the cumulative probability distribution of magnetars and young pulsars as references \citep{Ding_2024}.}
    \label{fig:hist}
\end{figure}

Half a century has passed since the discovery of the first pulsar in 1967 \citep{1968Natur.217..709H}. The number of known neutron stars is around 3,000, most of which are compiled in the Australia Telescope National Facility (ATNF) Parkes Pulsar Catalogue \citep{2005AJ....129.1993M} and its subsequent updates. Recent large-scale surveys with the Five-hundred-meter Aperture Spherical radio Telescope (FAST) have substantially increased this population, yielding hundreds of newly discovered pulsars across the Galactic plane \citep[e.g.,][]{Xue_2023, Han_2025}.

These neutron stars are usually classified on the parameter space between the spin (the period $P$ second) and spin derivative ($\dot{P}$ second/second), also known as the $P-\dot{P}$ diagram (Fig.\ref{fig:hist} left). While rotation-powered, ordinary pulsars form a majority of the distribution in the diagram, there is a class of neutron stars that possess the largest $P$ (typically above $1$~second) and the largest $\dot{P}$ (typically above $10^{-13}$~second/second). This class of neutron stars are normally found by $\gamma$-ray and X-ray observations as soft-$\gamma$ ray repeaters (SGRs) and anomalous X-ray pulsars (AXPs), respectively. The traditional formalizm, where the energy loss of magnetic dipole radiation is recovered by the spin energy (spin-down luminosity), implies that those neutron stars possesses extremely strong magnetic fields ($>10^{13}$~G as the dipole field equivalent), which govern their radiative processes. Therefore, they are called ``magnetars" \citep[see review by][]{2017ARA&A..55..261K}. The formalism also implies that mangetars are relatively young ($<1$~Myr as the characteristic age, $\tau_c = P/2\dot{P}$).

Little progress has been made in determining whether the strongest magnetic fields of magnetars are acquired or inherent. Through the gravitational collapse of the progenitor massive star into a neutron star, it is expected that the magnetic field is compressed and amplified. However, simple flux conservation during the collapse from a radius of $\sim 10^7$~km to 10~km can only account for magnetic field strengths comparable to those estimated for ordinary pulsars, and it is difficult to reach the extreme magnetic field strengths expected for magnetars. Therefore, some additional amplification mechanism is required. Significant dynamo amplification of magnetic fields is expected to occur in explosion mechanisms \citep{1992ApJ...392L...9D, 1993ApJ...408..194T} such as the Standing Accretion Shock Instability \citep[SASI; e.g.,][]{2003ApJ...584..971B, Foglizzo_2006}, which exhibit instable oscillation. Magnetars are thus expected to obtain larger kinetic energy as kick velocity compared to ordinal pulsars. Nevertheless, based on the currently limited sample, magnetars do not appear to exhibit peculiarly high transverse velocities (Fig.\ref{fig:hist} right). 

Studying magnetar activity is significant for understanding the physics of extremely strong magnetic fields, and it may offer important insights into the origin of such fields. For instance, glitches, which is sudden changes in the spin frequency, and outbursts, during which their X-ray and $\gamma$-ray luminosities increase dramatically, are often discussed with magnetic fields \citep[see review by][]{2017ARA&A..55..261K}. Interestingly, although magnetars are typically radio-quiet, several sources have been observed to emit transient pulsed radio emission following X-ray outbursts. The discovery of such radio activity is relatively recent, having emerged over the past two decades, and the underlying emission mechanism remains poorly understood. As of September 2025, there have been 7 radio outburst events from 5 magnetars. The prototypical example is XTE~J1810$-$197, which experienced radio outbursts twice (see Sec.2 for details).

Magnetar activity has also been highlighted as a promising candidate for the origin of fast radio bursts (FRBs). A major breakthrough occurred with the detection of FRB~200428 from the Galactic magnetar SGR~1935+2154, which established a direct observational connection between magnetars and FRBs \citep[e.g.,][]{2020Natur.587...54C, 2020Natur.587...59B}. Moreover, the recently discovered class of Long Periodic Transients (LPTs), which exhibit extraordinarily long periodicity, shows radio emission that is difficult to explain solely by rotational energy loss \citep[e.g.,][]{Rea_2022, 2023Natur.619..487H}. The energetics of these sources suggest that magnetic energy could be a dominant power source, and some of LPTs are thought to be evolved or aged magnetars.

\subsection{Importance of Magnetar Astrometry}

In order to solve various outstanding problems related to magnetars, precise astrometry of magnetars is a crucial breakthrough. Astrometry serves two main purposes. The first is to measure the proper motion of magnetars and examine whether they possess larger transverse velocities than ordinary pulsars. This will allow us to test the consistency with magnetic-field amplification mechanisms such as the SASI \citep[e.g.,][]{Foglizzo_2006, Burrows_2007}. The second purpose is to associate magnetars with their host supernova remnants (SNRs) in order to identify the characteristic properties of supernovae that give rise to magnetars. The characteristics contain SNR's morphology, especially an alignment of the SNR's major/minor axes to the motion of magnetars. Because the deceleration of a neutron star due to interaction with interstellar gas is estimated to be negligible \citep{1977MNRAS.180..717W}, it is possible to infer its birth position if its current three-dimensional position, velocity, and age are known . 

The age is, however, an uncertain parameter for both neutron stars and their associated SNRs. For neutron stars, the characteristic age $\tau_{\rm c}=P/2\dot{P}$ is often used as a proxy for the true age, although it can differ significantly depending on the braking history. For SNRs, the age is commonly estimated from the remnant size using theoretical expansion models. The diameter $D$ of a SNR can be estimated to be \citep[see][for a review]{2008ARA&A..46...89R}, $D \sim 2~{\rm pc} (t/100~{\rm yr})$ for young remnants in the initial free-expansion phase, and $D \sim 25~{\rm pc} (t/10~{\rm kyr})^{2/5}$ for the later adiabatic (Sedov-Taylor) phase, assuming typical values (the energy $10^{51}$~erg, the mass $1 M_\odot$, the density $1~{\rm cm}^{-3}$). The latter relation is referred to derive the Sedov time, $\tau_{\rm Sedov}$, as the age of the SNR.

SNRs should be observable in X-rays or other wavelengths if they are not too distant and the age is not too old ($< 1$~Myr) \citep{2008ARA&A..46...89R}. Nevertheless, the associated SNRs remain unidentified for nearly half of young neutron stars and magnetars which have $\tau_{\rm c} \sim 10$~kyr \citep{2019RPPh...82j6901E}. This is rather puzzling. Given a typical speed of 200~km/s, the time for the neutron star to escape a Sedov-phase SNR is $\sim 210$~kyr (assuming the SNR's center of mass is stationary), which implies that a neutron star with an age of $\sim 10$~kyr should still reside within its SNR. Possible explanations include: (1) some young neutron stars have kick velocities significantly higher than typical, or (2) their true ages are substantially older than the characteristic ages. Precise measurements of distance and proper motion are key to solving this puzzle.

\section{Previous works}

\begin{figure}[btp]
    \centering
	\includegraphics[width=\linewidth]{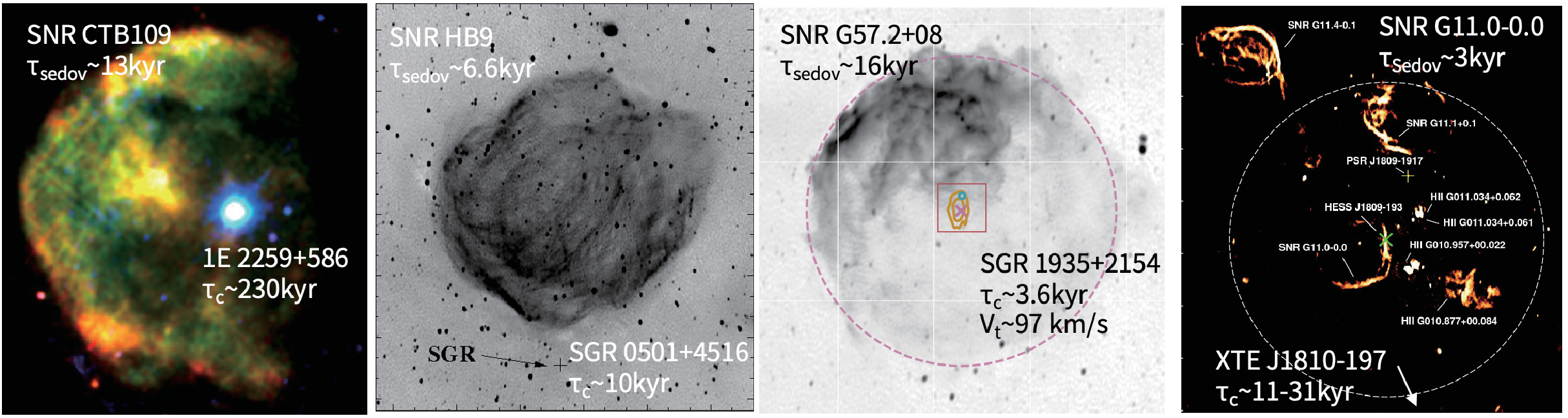}
    \caption{Images from left to right are SNR CTB109 \citep{2004ApJ...617..322S}, SNR HB9 \citep{2008GCN..8149....1G}, SNR G57.2+08 \citep{2022ApJ...926..121L}, and G11.0-0.0 complex \citep{2016A&A...587A..71C}.
    }
    \label{fig:SNRimages}
\end{figure}

Figure~\ref{fig:SNRimages} shows some examples of SNRs and magnetars. Magnetar 1E 2259+586 is apparently inside the SNR CTB109. However, $\tau_{\rm c}$ is an order of magnitude larger that $\tau_{\rm Sedov}$, raising a doubt about identifying this SNR as the host of the magnetar. SGR 0501+4516 is located to the south of SNR HB9. Multi-epoch observations with the Chandra X-ray Observatory have constrained the proper motion of the magnetar to be less than 0.32 arcsec/year to south assuming a distance of 5 kpc \citep{2018ApJ...852...86M}. Given $\tau_{\rm c}\sim 10$~kyr, it could have moved up to 3200 arcsec over its lifetime. However, the angular separation between the magnetar and the center of HB9 is 4800 arcsec, meaning that the magnetar could not have reached the SNR's center within its characteristic age. The conclusion that the magnitude and direction of the proper motion rules out supernova remnant HB9 as the birth site was also supported by HST imaging \citep{2025A&A...696A.127C}. Galactic FRB Mangetar SGR 1935+2154 is apparently near the center of SNR G57.2+08. The HST 6-year monitoring observations of SGR 1935+2154 suggested a small transverse velocity of $97\pm 48$~km/s for a distance of 6.6~kpc. Conversely to 1E 2259+586, $\tau_{\rm Sedov}$ of SNR G57.2+08 is factor of 4 larger than $\tau_{\rm c}$ of SGR 1935+2154. The latest success of VLBI radio astrometry of a magnetar was reported by \citet{Ding_2024} for Swift J1818.0-1607. They found a rather small transverse velocity of 48 km/s, which seems to be the opposite of the idea that a magnetar is violently ejected by a supernova explosion as described in Sec. 1.2.

XTE J1810-197 is a magnetar that activated in 2006 and reactivated in December 2018. Transient radio pulsations were first detected following its 2003 X-ray outburst \citep{2005ApJ...632L..29H,Camilo_2006}. The radio emission gradually faded and became undetectable by 2008. In multi-epoch VLBA observations conducted in 2006, the annual parallax could not be measured, but a proper motion of $212$~km/s toward the southwest was estimated assuming a distance of 3.3 kpc \citep{2007ApJ...662.1198H}. To trace this magnetar back to SNR G11.2-0.3, which lies to its northeast, would require 1,200 years in Galactic longitude and 7,500 years in Galactic latitude, resulting in inconsistent ages. $\tau_{\rm c}$ is estimated to be 31 kyr \citep{2019MNRAS.483.3832P} which also disagrees. Furthermore, G11.2-0.3 is estimated to be 1,633 years old based on ancient Chinese records, again inconsistent. The source underwent another outburst in 2018, accompanied by the reappearance of radio pulsations \citep{2018ATel12285....1D, 2018ATel12288....1L, 2018ATel12284....1L, Gotthelf_2019}. Over a 1.3-year, \citet{2020MNRAS.498.3736D} conducted 14-epoch VLBA observations of XTE J1810-197 at 5.7 GHz, measuring a distance of $2.5_{-0.3}^{0.4}$~kpc and a transverse velocity of $198_{-23}^{29}$~km/s. Tracing its motion back over 70~kyr would place it approximately 1 arcminute east of another faint SNR G11.0-0.0 (Fig.\ref{fig:SNRimages}). However, $\tau_{\rm Sedov}\sim 3$~kyr is inconsistent with $\tau_{\rm c}$ by an order of magnitude. While the uncertainty in the characteristic age is acknowledged, the roughly 20\% uncertainty in the velocity measurement also remains a significant issue.

Regarding parallax measurements, experience from pulsar VLBI astrometry provides important guidance for magnetar observations. A large VLBA project called PSR$\pi$ was carried out for pulsars. PSR$\pi$ conducted 2--4 monitoring epochs per year and determined annual parallaxes based on 8--10 total epochs. The targets were pulsars with flux densities between 0.6 and 30~mJy in the L band (except PSR B0329+54 at 200~mJy). Using the VLBA L-band at 21~cm (1.4~GHz) with an angular resolution of 5~mas, detections with $S/N= 5-10$ yielded final velocity uncertainties of $20-50~{\rm km/s}$. The nominal astrometric accuracy reported by PSR$\pi$ is 40~$\mu$as in position and 40 $\mu$as/yr in proper motion. In the first phase (2011-2013), observations were conducted for 60 pulsars, and in the second phase (2015-2018), 42 millisecond pulsars were observed. Annual parallax measurements were performed over eight epochs using in-beam calibrators. Such observational strategy and astrometric precision demonstrated by PSR$\pi$ provide a valuable framework for magnetar studies. Radio-emitting magnetars typically exhibit flux densities comparable to those of the pulsars targeted in the survey, suggesting that similar VLBI campaigns could achieve sub-100 $\mu$as astrometric accuracy (see also Sec 3). Such precision would enable reliable measurements of magnetar distances and transverse velocities (Sec 3), allowing their birth sites to be traced back and their physical associations with supernova remnants to be tested.

Note that astrometry conducted at L-band like PSR$\pi$ often adopts in-beam phase calibration, which enables high-precision measurements. In contrast, the probability of finding suitable in-beam calibrators is significantly reduced when we conduct astrometry at higher observing frequencies. We then apply phase-referencing with interpolation as the primary technique, making astrometry relatively worse. In contrast, the effect of interstellar medium (ISM) scattering decreases toward higher frequencies, making astrometry relatively better. Therefore, the optimal observing frequency should be selected by balancing the trade-off between phase calibration accuracy and ISM scattering effects (see Sec. 3.3).

Some SNRs exhibit structures called ``ears", which are thought to have been formed by jet-like outflows during the supernova explosion. There are studies that have investigated the correlation between the direction of these ears and the kick velocity of neutron stars found within the SNR. For example, \citet{2018ApJ...855...82B} argued that there is no correlation between the direction of ``ears" and the motion of its associated neutron star. In the case of the semicircular supernova remnant CTB 109 (1E 2259+586), the neutron star appears to be moving toward the direction where the semicircular shell is missing, suggesting a possible connection. However, a molecular cloud is also present in that direction, so the alignment could merely be coincidental.

\section{Magnetar Astrometry with SKA-VLBI}

\subsection{Cross Identification and angular resolution}

One reasonable method for cross-identifying a magnetar with the SNR that created it is to trace the neutron star's motion backward in time, using its position, velocity, and age. There are two possible cases for such cross-identification. The first case occurs when only one supernova remnant is found near the estimated region. In this case, even if the remnant lies outside the estimated region, it is still reasonable to regard it as the progenitor remnant if no other remnants are present (``weak cross-identification"). The second case occurs when the estimated region lies entirely within a SNR. In this case, it is reasonable to regard that SNR as the progenitor (``strong cross-identification").

In attempting to cross-identify magnetars with SNRs, it is valuable to understand how SNRs are distributed in the sky. More than 300 supernova remnants are dispersed throughout the Milky Way. Below, three representative SNR catalogs are introduced.
\paragraph{Green SNR Catalog (http://www.mrao.cam.ac.uk/surveys/snrs/)} This catalog contains 294 objects, of which 143 are located at declinations below $-20^\circ$. It provides information such as right ascension, declination, angular size, type (Shell, Filled-center, Composite), flux density at 1 GHz, and spectral index.
\paragraph{MOST SNR Catalog (http://www.physics.usyd.edu.au/astrop/wg96cat/)} This catalog lists 74 objects in the southern sky and provides flux density information at 843 MHz.
\paragraph{Chandra SNR Catalog (https://hea-www.harvard.edu/ChandraSNR/)} This catalog includes 129 objects and provides right ascension, declination, angular size, distances derived from the literature, and X-ray flux information.

The apparent angular sizes of SNRs are typically distributed between 1 and 30 arcminutes. To perform a meaningful cross-identification through a look-back analysis, the positional uncertainty of the magnetar must be small enough that the corresponding search region is comparable in size to a typical SNR. The average surface density of SNRs along the Galactic plane is approximately one per $30'\times 30'$ area in the central quadrants toward the Galactic center. We consider that a weak cross-identification requires only one SNR to fall within the estimated region, so that a practical benchmark is about one-third of the mean separation between SNRs, $\sim 10' \sim 3~{\rm pc}~(d/{\rm kpc})$, where d is the distance from the observer. Assuming a typical age of 10~kyr, this translates to a required proper-motion precision of about $60~{\rm mas/yr}~(d/{\rm kpc})^{-1}$ for the weak cross-identification in the central quadrants. A similar discussion is available for the anti-center quadrants, where the SNR density is one per $10^\circ \times 10^\circ$ area, so that $\sim 3^\circ \sim 50~{\rm pc}~(d/{\rm kpc})$. With a age of 10~kyr, it translates into the required proper-motion precision of $1~{\rm arcsec/yr}~(d/{\rm kpc})^{-1}$.

For the strong cross-identification, we consider that the extrapolated birth position of the magnetar fall within the central 10\% of the SNR, the corresponding spatial accuracy must be within $2.5~{\rm pc}~(t/10~{\rm kyr})^{2.5}$. This implies a required proper-motion precision of about $50~{\rm mas/yr}~(d/{\rm kpc})^{-1}$; roughly the same order of accuracy as that needed for weak cross-identification in the central quadrants.

\subsection{Parallax Measurement}

Distance measurement is essential for converting observed quantities (such as mJy) into physical quantities (such as erg/s), thereby enabling distance-independent comparisons. Conventional estimates based on electron column densities in radio or X-ray observations often carry uncertainties of up to a factor of a few. From this perspective, distance determination through annual parallax measurements is an important approach. By definition, the annual parallax is $1~{\rm mas/yr}~(d/{\rm kpc})^{-1}$. For instance, if the annual parallax can be measured with an accuracy of 50 $\mu$as, the distance to an object at 1 kpc (parallax = 1 mas) can be determined with a 5 \% uncertainty, which will be enough for cross-identification. However, even with the same accuracy of 50 $\mu$as, the distance error becomes 40--60 \% (or the distance uncertainty ranges 5.71--13.3 kpc) for an object at 8 kpc. Therefore, targeting nearby magnetars is strongly recommended.

The positional accuracy is primarily determined by the statistical position error governed by thermal noise, which is proportional to the angular resolution and inversely proportional to the signal-to-noise ratio (S/N). Thus, higher S/N detections allow the source centroid to be determined with a precision significantly smaller than the nominal angular resolution. The accuracy also depends on the number of epochs. When only a small number of epochs are available, the fitted parameters (position, proper motion, and parallax) remain strongly correlated, and the uncertainty decreases approximately inversely with the number of epochs. Once the time sampling becomes sufficient to separate these parameters, the uncertainty decreases approximately as the inverse square root of the number of epochs, as seen for pulsar astrometry in which many pulsar pulse signals can be easily added together. Actually, the positional accuracy achieved in the PSR$\pi$ project can be estimated as approximately, (angular resolution 5 mas) / (S/N 10) / (10 epochs) $\sim$~50~$\mu$as, which is broadly consistent with the nominal astrometric precision reported by the survey, i.e. 40~$\mu$as in position and 40 $\mu$as/yr in proper motion.

It should be noted that, while pulsar astrometry often benefits from pulse gating, which allows many pulses to be coherently integrated to improve the signal-to-noise ratio, radio magnetars also possess stable spin periods and their signals can in principle be folded in the same manner. However, magnetar radio emission frequently exhibits significant variability in pulse morphology and flux density, which may reduce the effectiveness of pulse gating compared to ordinary pulsars. In addition, the episodic nature of magnetar radio activity can complicate long-term astrometric campaigns required for parallax measurements if the emission becomes weak or disappears during part of the observing interval.

If the radial velocity could also be determined in addition to the proper motion, the three-dimensional velocity could be derived, which would further help identify the origin SNR. However, measuring the radial velocity is challenging for the following reasons. First, line emission have never been detected from magnetars, so the radial velocity cannot be estimated via redshift. Second, for a typical neutron star velocity of 200 km/s, the displacement along the line of sight over 10 kyr is only $\sim 2$~pc, making it difficult to detect any change in parallax due to depth motion. Furthermore, in pulsar timing, there is a degeneracy between the pulsar's radial velocity and its rotation period; if both are unknown, there is no way to separate them. Nevertheless, for ordinary pulsars, assuming an isotropic velocity distribution, studies have estimated the 3D velocity distribution from the 1D velocity along a particular axis and the 2D proper motion on the sky \citep[see e.g.,][]{2005MNRAS.360..974H, 2017A&A...608A..57V}.

\subsection{Sensitivity}

The required sensitivity is determined by the brightness of the target source. In the 2020 outburst of SGR 1935+2154, strong bursts were detected with fluences of 1-2 kJy ms at 0.4-0.8 GHz and a peak flux density of 1.5 MJy at 1.4 GHz. Two days later, the flux density was 30 mJy (or 60 mJy ms) at 1.25 GHz. In the 2006 outburst of XTE 1810-197, the flux densities were 22.5 mJy at 4.74 GHz and 14.9 mJy at 8.35 GHz, corresponding to a spectral index of -0.73. During the 2020 outburst of Swift J1818.0-1607, 29 burst events were detected in the 4-8 GHz band, with flux densities ranging from 21 mJy to 500 mJy. Assuming the same spectral index (-0.73) for the 2006 XTE 1810-197 event, the expected flux density at 22 GHz would be 7.4 mJy. In the 2018 outburst of XTE 1810-197, a flux density of about 20 mJy was observed at 22 GHz. Outbursts are therefore readily detectable with a sensitivity of $\sigma \sim 1~{\rm mJy}$. A flat spectral index means that, compared with pulsars, magnetars remain bright even at high radio frequencies at which we have a higher angular resolution and higher position accuracy in general.

The SKA station refers to the phased-up central part of the SKA1-MID array, which forms a single VLBI beam. The subarray used for phase-up corresponds to the ``core" region, defined as within a radius of 3--10 km. Out of the total 197 antennas of SKA AA4, about 154 (approximately 78\%) are planned to be located within this core. Among them, 90 are SKA dishes with an effective diameter of 15 m, and the remaining 64 are MeerKAT dishes with an effective diameter of 13.5 m. In the SKA's Central Signal Processor (CSP), both types of antennas can be combined coherently to form a single beam. Considering the 90 SKA dishes and 64 MeerKAT dishes, and assuming that the MeerKAT dishes have 70\% of the SEFD of the SKA dishes (i.e., equivalent to 45 SKA dishes), and further assuming—based on phased-array performance from the VLA—that the SEFD degrades to 70\% after beamforming, we can roughly approximate the following relation for frequencies above 1 GHz: ${\rm SEFD_{SKA-MID-Core}} \sim 5~{\rm Jy} \exp [0.075(\nu/{\rm GHz})]$. In the AA* configuration, using about 60 SKA dishes results in an SEFD of approximately 6.5 Jy. Those values are a 1--2 order of magnitude improvement in sensitivity compared to those of existing VLBI stations around the world.

\subsection{Partner Stations and Use Cases}

To achieve a strong cross-identification, the proper motion of the neutron star must be determined with an uncertainty better than $50~{\rm mas/yr}~(d/{\rm kpc})^{-1}$. To achieve 50 mas at a distance of 1 kpc, a baseline of approximately 2300 km is required at 700 MHz. With the same baseline, observations at 5.6 GHz would achieve the same resolution for a source at 8 kpc, allowing measurements for magnetars near the Galactic center. More specifically, at 12 GHz, the angular resolution will be 0.86 mas, 0.62 mas, and 0.57 mas for the baselines between SKA-Mid and, for example, Yebes 40m (distance 7700~km), Tianma 65m (10600~km), and Mizusawa 20m (11600~km), respectively \citep[See also][for expected specifications in detail]{Rioja01.2026.SKA}. These resolution is comparable to or better than the monthly movement of the magnetar on the plane of the sky with a nominal tangential velocity of 200~km/s, 1.7 mas/month for d=2 kpc (local arm), 0.85 mas/month for d=4 kpc (inter-arm), and 0.45 mas/month for d=8 kpc (galactic center). To resolve the motion, we obtain the required duration of at least the two observing epochs to measure the proper motion. As for the case of SKA-Low, the yearly movement could be resolved by the resolution, 38.4 mas, 34.8 mas, and 19.3 mas for the baselines between SKA-Low and, for example, GMRT (6900~km), Iitate (7600~km), and LOFAR (13700~km), respectively \citep[See also][for expected specifications in detail]{Kobayashi01.2026.SKA}.

As for the case of parallax measurement, motivated by previous PSR$\pi$ project (See Sec 3.2), we require astrometric precision of 50~$\mu$as in position and 50 $\mu$as/yr in velocity. Remember the relation 50~$\mu$as $\sim$ (angular resolution 5 mas) / (S/N 10) / (10 epochs), SKA-VLBI with the subprime sensitivity described above will expand the horizon of parallax measurement of magnetars, from currently $\sim 2$~kpc or so to the Milky Way overall based on the improvement of the signal to noise. International baselines described above further improve the angular resolution, so that they will expand the distance horizon too.

We expect that AA4 improves sensitivity by about 30\% compared to AA*. This means that, at the same signal-to-noise ratio, it can detect objects at a distance about 70\% farther away, expanding the volume of the research area by a factor of $\sim 5$.

\section{By-products from monitoring observations}

VLBI astrometry of magnetars is conducted through observations spanning multiple epochs. The data obtained from such monitoring observations is useful not only for astrometry but also examining the physics under extremely strong magnetic fields ($10^{14}~{\rm G}\sim 580~{\rm keV}$). They enable the investigation of quantum electrodynamics (QED) and quantum chromodynamics (QCD) processes. Here we briefly discuss such a ``by-product" from monitoring observations of magnetars.

\subsection{Why are magnetars usually radio-quiet?}

The physical mechanism of neutron-star radio emission remains only partly understood, even decades after discovery. A key clue is the coherent emission with brightness temperatures up to $T \sim 10^{20}$ K, implying particle acceleration by strong magnetospheric voltages, curvature radiation along magnetic field lines, high-energy photon–induced pair production, and cascading. A major distinction between pulsars and magnetars is that in magnetars the magnetic field is so strong that the electron gyroradius becomes comparable to or smaller than the Compton wavelength. This threshold, the quantum critical field $B_{\rm cr}$, defines the regime where QED processes dominate.

One proposed explanation for the general lack of radio emission from magnetars is that magnetic photon splitting may suppress pair-production cascades \citep[see][for a review]{2008AIPC.1051...53B}. This QED process splits a single photon into two lower-energy photons. During quiescence, strong magnetic fields may suppress radio emission, while this suppression can relax during outbursts. Other possibilities include strongly beamed emission away from the observer or location in dense interstellar material that heavily scatters radio waves. Observing more magnetar outbursts or identifying neutron stars without long-term outbursts may help determine the threshold for steady radio pulses, testing the photon-splitting model. Investigating whether radio and X-ray emissions are correlated or anti-correlated is also important. The intensity and spectra of thermal and nonthermal X-ray components, and their relation to radio emission, could provide key insights.

Detecting wideband radio pulses with the SKA and measuring their spectra is crucial. It is important to determine whether the spectrum follows a single power law, broken power laws, or a smooth curve with varying index, thereby constraining emission models. Identifying low- and high-frequency cutoffs will define where coherent emission cannot occur or escape, constraining the physical conditions of the emission and propagation regions.

\subsection{Link with Fast Radio Bursts (FRBs) and Giant Radio Pulses (GRPs)}

The spectral index of radio emission during outbursts is much flatter ($\alpha \sim 0$) than that of rotation-powered pulsars (RPP) ($\alpha \sim -1.6$), with pulse-to-pulse variations observed \citep[e.g.,][]{Camilo_2006, 2015MNRAS.451L..50T}. Similar shallow, sporadic radio emission is also seen in some FRBs \citep[e.g., FRB~121102:][]{2016Natur.531..202S, 2019ApJ...876L..23H}, suggesting a possible connection, though its generality is unclear. In contrast, radio outbursts from Swift J1818.0-1607 (0.7–0.9 mJy at 1.4–1.5 GHz) show a steep spectral index \citep[$\alpha \lesssim -1.8$:][]{2020MNRAS.498.6044C, 2020ApJ...896L..37L} immediately after the outburst, indicating increasing complexity.

The intensity distribution of steady pulses from ordinary pulsars follows a log-normal distribution \citep{2001ApJ...563L..65C, 2012MNRAS.423.1351B}, whereas magnetar outbursts often follow a power law (with some log-normal exceptions) \citep[e.g.,][]{10.1111/j.1365-2966.2008.14260.x}. It remains necessary to test whether this trend is universal. Similar power-law behavior is seen in GRPs from the Crab pulsar, suggesting magnetar outbursts are statistically similar to GRPs \citep{1995ApJ...453..433L}.

The frequency distribution reflects the charged-particle population and is key to understanding the emission mechanism. Coherent emission is intrinsically narrowband, so a continuous spectrum likely arises from superposition across multiple regions. In ordinary pulsars, frequency-dependent pulse phase shifts support this view. If the frequency distribution varies with observing frequency, it can constrain environments favorable for coherent emission (e.g., distance from the neutron star or magnetic latitude) or reveal propagation efficiency effects. It is also important to examine the frequency dependence of outburst decay; in past cases, higher frequencies decay faster \citep{2021PASJ...73.1563E}.

Building a statistical sample to identify the highest-energy pulses is valuable, as it constrains the energy, especially free energy, in the radio-emitting region. This is also relevant for testing the magnetar FRB hypothesis. SGR 1935+2154 was observed with radio fluences of several kJy ms \citep[CHIME, 400--800 MHz:][]{2020Natur.587...54C} and 1.5 MJy ms \citep[STARE2, 1.4 GHz:][]{2020Natur.587...59B}, attracting attention as an FRB candidate.

\subsection{Exploration of Axion and Axion-Like Particles (ALPs)}

QCD, describing the strong interaction, permits a charge-parity (CP)-violating term in the Lagrangian. However, no CP violation has been observed in purely strong interactions, leading to the strong CP problem. The axion is a hypothetical pseudo-scalar particle proposed in Peccei-Quinn theory as a solution, emerging as a Nambu-Goldstone boson from spontaneous symmetry breaking. The XENON1T experiment \citep{2020PhRvD.102g2004A} reported an unexpected 3.5$\sigma$ excess (99.98\%), drawing attention as a possible axion signal.

Let $\omega$ be the photon energy ($\hbar = 1$) and $B_{||}$ the magnetic-field component parallel to the photon polarization plane. The eigenvalues of photon-axion mixed states are $m_\pm^2 = \pm gB_{||}\omega$, with refractive indices $n_\pm = 1 - m_\pm^2/2\omega^2$. Thus, in an axion and magnetic field, photons split into two polarization modes, producing wavelength-independent polarization rotation from phase-velocity differences. Recent CMB observations provide evidence for axion-like particles (ALPs) at non-negligible significance \citep{2020PTEP.2020j3E02M, 2022PhRvD.106f3503E}, and they are discussed as dark matter or dark energy candidates \citep{2025arXiv250406709N}. From polarization-angle dispersion of PSR J0437-4715 over 393 data points across 1609 days, a constraint of $g \lesssim 10^{-12} {\rm GeV}^{-1}(m_a/10^{-21}~{\rm eV})$ has been derived \citep{2019PhRvD.100f3515C}. Distinguishing whether pulse arrival times align or shift across frequencies, and whether pulse widths match or differ, can probe the emission region and underlying physics.

A shift in pulse arrival time, $\delta \phi \sim 0.01(g/10^{-12}{\rm GeV}^{-1})(B/10^{15}{\rm G})(\omega/2{\rm GHz})$~rad, is also expected as a second-order effect from group-velocity changes
\citep{2011IJMPE..20..100G}. This shift could be resolved with a temporal resolution of $\sim 1/10^4$ for the spin period. To identify the characteristic splitting width, it is important to examine the frequency distribution of pulse-peak phases. The pulse beam may remain single while only the phase shifts (by half the splitting phase difference). Since this shift is undetectable at a single frequency, simultaneous multi-frequency observations are essential. In strong magnetic fields, the Primakoff effect predicts photon-axion conversion. If some axions escape, the radio flux may decrease. Using the frequency dependence of this conversion may enable axion detection through the observed spectral shape.

\section{Summary}

The birth and evolution of neutron stars, as well as the mechanism of supernova explosions, are among the key topics in astronomy and astrophysics. Neutron stars exist in various classes, but among them, those with rapidly decelerating spin periods pose many mysteries. This class, magnetar, is believed to possess extremely strong magnetic fields, yet it is unknown why their magnetic field strengths are significantly higher than those of other neutron stars. Moreover, despite being estimated to be relatively young, many examples lack an identifiable parent supernova remnant. Magnetars often exhibit explosive brightening events, but the causes of these outbursts remain unclear. There is also the fundamental question of how quantum chromodynamics—including axions—and quantum electrodynamics phenomena such as photon splitting manifest themselves in the “extreme magnetic field environments,” which are both inaccessible on Earth and remain unexplored.

We will conduct long-term monitoring observations magnetar radio outbursts. To identify the parent SNR, we will track proper motions with a precision of 50~mas/yr or~4 mas/month and determine distances via annual parallax measurements with an accuracy of 50 $\mu$as, enabling us to associate the magnetar with its SNR and estimate its age. We will also establish the broadband radio emission spectra and pulse profiles of this population, leading to significant progress in understanding the emission mechanisms. Furthermore, by exploiting statistics from a large number of pulses, we will obtain world-leading, independent constraints on the axion mass and coupling constant.

Finally, magnetar radio outburst is generally bright so that a sensitivity of $\sigma \sim 1~{\rm mJy}$ level is enough. A relatively-flat spectral index allows us to detect the outburst even the frequency at Band 5 and beyond. 

\bibliographystyle{abbrvnat-maxbibnames4}
\bibliography{chapter}

\end{document}